\documentclass[preprintnumbers, floatfix,letterpaper,aps,prd,epsfig,nofootinbib,
twocolumn
]{revtex4-1}
\usepackage{bm,graphicx,dcolumn,epstopdf,epsf, latexsym,mathbbol, amssymb,amsmath,color,slashed, mathrsfs,mathcomp, simplewick}
\usepackage[center]{subfigure}
\usepackage{multirow}
\usepackage{makecell}
\usepackage[colorlinks,linkcolor=blue,citecolor=blue,urlcolor=blue]{hyperref}

\begin{document}
	\allowdisplaybreaks
	\newcommand{\bq}{\begin{equation}}
	\newcommand{\eq}{\end{equation}}
	\newcommand{\bqn}{\begin{eqnarray}}
	\newcommand{\eqn}{\end{eqnarray}}
	\newcommand{\nb}{\nonumber}
	\newcommand{\lb}{\label}
	\newcommand{\f}{\frac}
	\newcommand{\p}{\partial}
	\newcommand{\PRL}{Phys. Rev. Lett.}
	\newcommand{\PLB}{Phys. Lett. B}
	\newcommand{\PRD}{Phys. Rev. D}
	\newcommand{\CQG}{Class. Quantum Grav.}
	\newcommand{\JCAP}{J. Cosmol. Astropart. Phys.}
	\newcommand{\JHEP}{J. High. Energy. Phys.}
	\newcommand{\red}{\textcolor{black}}
	
\title{Observational tests of quantum extension of Schwarzschild spacetime in loop quantum gravity with  stars in the galactic center}

\author{Jian-Ming Yan${}^{a, b}$}
\email{yanjm@zjut.edu.cn}

\author{Cheng Liu${}^{a, b}$}
\email{liuc09@sjtu.edu.cn}

\author{Tao Zhu${}^{a, b}$}
\email{zhut05@zjut.edu.cn; Corresponding author}
	
\author{Qiang Wu${}^{a, b}$}
	
\author{Anzhong Wang${}^{c}$}
	
\affiliation{${}^{a}$Institute for Theoretical Physics \& Cosmology, Zhejiang University of Technology, Hangzhou, 310023, China\\
${}^{b}$ United Center for Gravitational Wave Physics (UCGWP),  Zhejiang University of Technology, Hangzhou, 310023, China\\
${}^{c}$ GCAP-CASPER, Physics Department, Baylor University, Waco, Texas 76798-7316, USA}
	
\date{\today}
	
\begin{abstract}
In this paper, we use the publicly available observational data of 17 stellar stars orbiting Sgr A* to test the quantum extension of Schwarzschild spacetime in loop quantum gravity (LQG). For our purpose, we transform the geodesical evolution of a massive particle in the quantum-extended Schwarzschild black hole to the perturbed Kepler problem and calculate the effects of LQG on the pericentre advance of the stellar stars. With these effects, one is able to compare them with the publicly available astrometric and spectroscopic data of stellar stars in the galactic center. We perform Monte Carlo Markov Chain (MCMC) simulations to probe the possible LQG effects on the orbit of S-stars. No significant evidence of the quantum-extended Schwarzschild black hole from LQG is found. Among the posterior analyses of 17 S-stars, the result of S2 gives the strongest bound on the LQG parameter $A_\lambda$, which places an upper bound at 95\% confidence level on $A_\lambda$ to be $A_\lambda < 0.302$. 
	
\end{abstract}

\maketitle
	
\section{Introduction}

Over the past century, general relativity (GR), as a successful theory of gravity, has passed all the experimental and observational tests, such as the perihelion shift of astrophysical bodies \cite{testGR, Mercury}, the deflection of light \cite{VLBI_deflection}, the time dilation \cite{cassini}, the observations of binary pulsars \cite{pulsar1, pulsar2}, the black hole image \cite{Akiyama:2019fyp, Akiyama:2019eap}, and the detection of gravitational waves \cite{gws}. However, there are still some problems with the observations, especially at galactic and cosmological scales. The components of dark matter and dark energy are still not clear and waiting to be clarified.

Therefore, in some ways, one can say that GR is still an incomplete theory. For example, GR predicts the existence of spacetime singularities where the theory itself ceases to be valid \cite{hawking}. To solve the singularity problem, we are now pinning our hopes on some quantum effects. A major challenge in solving such problems is figuring out how to formulate a consistent quantum theory of gravity. In the past decades, several approaches towards quantum gravity have been developed, such as loop quantum gravity (LQG), canonical quantum gravity \cite{AOS18b}, string theory \cite{BBy18}, and Euclidean path integral \cite{ABP19, ADL20, Perez17}. Since none of these methods are complete in themselves, from a more phenomenological point of view, it is recommended to consider effective quantum corrections that may help correct GR when considering systems with large curvature scales. This modified theory of gravity can be viewed as valid, or as a semi-classical approximation to the unknown full quantum theory of gravity. Among the various approaches, in this paper, we will consider the LQG scenario and focus on the effective black hole model in this approach. 

As a candidate theory of quantum gravity, LQG is a non-perturbative and background-independent approach to quantizing gravity which is based on the Ashtekar-Barbero variables, namely the $SU(2)$ connection $A_a^i$ and the densitized triads $E_i^a$ \cite{rovelli-book,Thiemann:2007pyv}. Concerning this conjugate pair, a classically background-independent $*$-algebra, known as the holonomy-flux algebra, is constructed and the classical Hamiltonian and diffeomorphism constraints are expressed in terms of the holonomies and the fluxes. A quantum representation of the holonomy-flux algebra leads to LQG. One of the key predictions of LQG concerns the discrete spectra of the volume and area operators which shed light on the intrinsic properties of quantum geometry \cite{Ashtekar:1996eg,Ashtekar:1997fb}. In particular,  the lowest non-zero eigenvalue of the area operator plays a vital role in a self-consistent formulation of standard loop quantum cosmology in a spatially-flat Friedmann-Lemaître-Robertson-Walker(FLRW) universe \cite{Ashtekar:2006wn}. Although the Hamiltonian constraint operator can be written explicitly in terms of the holonomy and volume operators, due to its complexity, the dynamics of full LQG are still left unraveled. In LQG, Dirac observables are physical observables that are invariant under the gauge transformations generated by the Hamiltonian constraint \cite{Thiemann:2005zg}. These observables are significant because they are the only ones that have a well-defined evolution in time in the quantum theory. They represent measurable quantities that are independent of the choice of the reference frame, which is a fundamental requirement in any theory of gravity. Examples of Dirac observables in LQG include the area and volume operators of spatial sections and the holonomies of connections along curves. By constructing and studying the properties of Dirac observables, LQG provides a framework for understanding the quantum nature of spacetime and its geometry. The general idea behind building efficient models in LQG is that one expects some non-perturbative quantum geometry corrections, which will modify the Einstein equations. One would expect to write down such semi-classical corrections that, when considering the full quantum theory, the expected value of the operator would appear in some well-defined semi-classical state  \cite{Rovelli18}.


With the above motivations, the effective models for describing black holes in the framework of LQG have attracted a lot of attention recently \cite{Hossenfelder:2012tc, Sahu:2015dea, AOS18a, Cruz:2015bcj, Caravelli:2010ff, AzregAinou:2011fq, Yan:2022fkr, Ongole:2022rqi, Zhu:2020tcf, Liu:2020ola, Liu:2022qiz, Fu:2021fxn, Fu:2022afk}. In practice, for effective versions of a black hole in LQG, one typically uses a polymerlike quantization inspired by LQG, in which the quantum theory of black hole is achieved by replacing the canonical variables $(b, c)$ in the phase space of the black hole spacetime with their regularized counterparts, $b \to \frac{sin(\delta_b b)}{\delta_b}$ and $c \to \frac{\sin(\delta_c c)}{\delta_c}$ \cite{Ashtekar20}. Here $\delta_b$ and $\delta_c$ are two quantum polymeric parameters that control the relevant scales of the quantum effects of LQG. However, as we do not yet have a complete quantum gravity, a full picture for determining $\delta_b$ and $\delta_c$ is still lacking. 

There are several different choices of $\delta_b$ and $\delta_c$, which can be divided into three classes \cite{Gan:2020dkb}: the $\mu_0$-scheme \cite{Ashtekar:2005qt, Modesto:2008im, Assanioussi:2019twp, Modesto:2005zm, Corichi:2015xia}, $\bar \mu$-scheme \cite{ABP19, Boehmer:2007ket, Gan:2022oiy, Ongole:2022rqi, Gan:2022mle, Garcia-Quismondo:2022ler, Bodendorfer:2019xbp}, and the generalized $\bar \mu$-scheme \cite{AOS18b, AOS18a}. For $\mu_0$-scheme, the two polymetric parameters are treated as constants and the corresponding effective LQG black hole has been explored in \cite{Modesto:2008im, Modesto:2005zm}. In $\bar\mu$-scheme $\delta_b$ and $\delta_c$ are treated as phase space functions that depend on the canonical variables \cite{ABP19, Boehmer:2007ket}. And in the generalized $\bar\mu$-scheme, $\delta_b$ and $\delta_c$ are considered as Dirac observables \cite{AOS18b, AOS18a}. Very recently, a quantum extension of the Schwarzschild black hole is constructed based on a specific $\bar \mu$-scheme \cite{LQG_BH1, LQG_BH2}. In such an effective quantum spacetime, similar to the case in loop quantum cosmology, the spacetime singularity of the classical Schwarzschild black hole can be replaced by a quantum bounce that connects the black hole region and the white hole region. \red{In this picture, the quantum effect is controlled by the parameter $A_\lambda$ which sensitively depends on $\delta_b$ and $\delta_c$, and its exact value in LQG has not been determined yet. It is interesting to see if the experiments or observations can lead to any bounds on it. } Based on this quantum-extended Schwarzschild black hole, a rotating spacetime with the LQG effects has been constructed by the Newman-Janis algorithm \cite{Brahma:2020eos}.

Naturally, one might wonder if there is any experiment or observations to test the LQG effects on this quantum-extended Schwarzschild black hole. In fact, many observational/phenomenological tests have been done to constrain the LQG effects on black holes. For example, in \cite{Papanikolaou:2023crz}, how quantum effects can influence primordial black hole formation within a quantum gravity framework has been discussed in detail. In addition, people have also tested the LQG black holes with the Event Horizon Telescope observations \cite{Islam:2022wck, Afrin:2022ztr} and constrain the parameter arises in LQG black hole with the observational data of M87* and Sgr A* \cite{KumarWalia:2022ddq, Yan:2022fkr, Vagnozzi:2022moj}. Some other phenomenological studies on testing LQG black holes can be found in \cite{Brahma:2020eos, Pugliese:2020ivz, Pawlowski:2014nfa, Vaid:2012pr, Barrau:2011md, Modesto:2009ve} and references therein.

The main purpose of this paper is to explore the LQG effects of the quantum-extended Schwarzschild black hole on the orbits of the stellar stars orbiting the supermassive black hole Sgr A* in the galactic center. We calculate the effects of LQG on the pericentre advance of the stellar stars and compare the orbits with the publicly available astrometric and spectroscopic data of stellar stars (S-stars) in the galactic center to obtain the constraints on the LQG parameter. For our purpose, we consider recent observations of 17 S-stars
\bqn
\{S1, S2, S4, S8, S9, S12, S13, S14, S17, \nb\\
 ~~~ S18, S19, S21, S24, S31, S38, S54, S55 \} \nb
\eqn
orbiting around Sgr A* to constrain the parameter $A_{\lambda}$ arising in the quantum-extended Schwarzschild spacetime in LQG. These stars orbiting around the black hole give us an opportunity to probe gravity in a strong field regime to test GR \cite{GRAVITY2, DellaMonica:2021xcf}. Also, these observations also allow us to constrain black holes in different gravitational theories, for example, see Refs.~\cite{DellaMonica:2021xcf, DAddio:2021smm, DellaMonica:2021fdr,deMartino:2021daj}. Not only that, more works have been done like testing the no-hair theorem \cite{nohair, nohair2}, and studying a black hole with dark energy interaction \cite{Benisty:2021cmq}. 

The plan for the rest of the paper is as follows. Section II provides a brief introduction to the quantum extension of the Schwarzschild spacetime in LQG, followed by an analysis of the geodesic equations for massive objects in this spacetime. Using these equations, we present a detailed derivation of the effects of the loop quantum parameter $A_\lambda$ on the pericenter advance of orbits. In Section III, we describe the dataset used in our Monte Carlo Markov Chain (MCMC) analysis, which includes the data of positions and velocity of 17 S-stars, and the orbital precession of S2. Section IV presents the upper bound on $A_\lambda$ obtained by comparing theoretical predictions of the quantum-extended Schwarzschild black hole in LQG with astrometric and spectroscopic data of stellar stars in the galactic center from an MCMC analysis. Finally, Section V summarizes our main results and provides some discussion. Additionally, in Section VI, we offer further insights and perspectives on our findings.

Throughout the paper, we use the units to $G=c=1$.

\section{LOOP QUANTUM BLACK HOLE}

In this section, we provide a brief introduction to the quantum-extended Schwarzschild spacetime in the framework of LQG and derive the equation of motion of a massive particle orbiting it. In order to calculate the effect of the LQG on the orbital precession of massive objects, we transform the geodesic equations of the massive particle into the perturbed Kepler problem in celestial mechanics.

\subsection{Quantum extension of the Schwarzschild spacetime in LQG\label{secrot}}
	
In this subsection, we introduce the effective quantum-extended Schwarzschild spacetime in LQG, which arises from a symmetry-reduced model of LQG corresponding to homogeneous spacetimes and is geodesically complete. The metric of this quantum-extended Schwarzschild spacetime is given by \cite{LQG_BH1, LQG_BH2}
\begin{equation}\label{1}
ds^2= - f(x)d \tau^2 + \frac{dx^2}{f(x)} + h^2(x)(d\theta^2+\sin^2\theta d\phi^2),
\end{equation}
where the metric functions $f(x)$ and $h(x)$ are given by
\begin{eqnarray}
f(x)&=&8A_\lambda M^2_B \Bigg(1-\sqrt{\frac{1}{2A_\lambda}}\frac{1}{\sqrt{1+x^2}}\Bigg)\frac{1+x^2}{h(x)^2},\\
h^2(x)&=&\frac{A_\lambda}{\sqrt{1+x^2}}\frac{M^2_B(x+\sqrt{1+x^2})^6+M^2_W}{(x+\sqrt{1+x^2})^3}. \lb{hx}
\end{eqnarray}
Here $M_B$ and $M_W$ correspond to two Dirac observables in the model, and $A_\lambda$ is a dimensionless parameter which is related to $M_B$ and $M_W$ via
\bqn
A_\lambda \equiv (\lambda_k/M_B M_W)^{2/3}/2,
\eqn
where $\lambda_k$ originates from holonomy modifications in LQG. Without loss of generality, in this paper, we focus on the most physical and meaningful case with $M_B=M_W=M$. It is convenient to define two new variables, $t$ and $y$, as
\bqn \lb{trans}
t=\frac{\tau}{\sqrt{8A_\lambda}M},~~~~ y=\sqrt{8A_\lambda}M x,
\eqn  
then metric (\ref{1}) can be rewritten as
\bqn
ds^2&=&-8A_\lambda M^2 f(y)dt^2+\frac{1}{8A_\lambda M^2 f(y)} dy^2 \nb\\
&& +h(y)^2 d\Omega^2.
\eqn
It is interesting to note that $h^2(y)$ represents the physical radius of the above spherical symmetric spacetime. In this way, one can rewrite the above metric by changing $dy \to dh(y)$. This can be achieved by writing out the asymptomatic form of the quantum-extended Schwarzschild metric as
\bqn \lb{7}
ds^2 &=&-\Bigg(1-\frac{2M}{h}+\frac{6A_\lambda M^2}{h^2}\Bigg)dt^2\nb\\
&&+\Bigg(1 + \frac{2M}{h}+\frac{4(1-2A_\lambda)M^2}{h^2} \Bigg)dh^2+h^2d\Omega^2. \nb\\
\eqn
Note that the metric functions in the above metric are expanded about $M=0$ to the next-to-leading order. It is easy to verify that the metric for the case $A_\lambda=0$ reduces to the Schwarzschild metric with a mass $M$ at the asymptotic region. 

\red{We would like to mention here that we only consider the static black hole spacetime and ignore the effects of the angular momentum of the black hole. We expect the observational effects in the stellar stars are expected to be very small. }

\subsection{Equations of motion of massive objects in the quantum-extended Schwarzschild spacetime}

Our purpose here is to study the motion of massive test particles in the quantum-extended Schwarzschild spacetime, which obeys time-like geodesics. It is convenient to transform the coordinates $(h, \theta, \phi)$ into the isotropic coordinates $(r, \theta, \phi)$, namely
\bqn
h=r\left(1+\frac{M}{2r}\right)^2.
\eqn
With the coordinates $(r, \theta, \phi)$, the metric of the quantum-extended Schwarzschild spacetime is written as
\bqn
ds^{2}=-f(r)dt^{2}+g(r)\Big[dr^{2}+r^{2}(d\theta^{2}+\sin^{2}\theta d\phi^{2})\Big],\lb{iso}
\eqn
where
\bqn
f(r)&=&\left(1-\frac{2M}{r}+\frac{2M^{2}(1+3A_{\lambda})}{r^{2}}\right),\\
g(r)&=&\left(1+\frac{2M}{r}-\frac{2M^{2}(-1+4A_{\lambda})}{r^{2}}\right).
\eqn

Let us now consider the motion of a massive object in quantum-extended spacetime. The massive object, if one ignores its self-gravitational effects, follows a time-like geodesic that reads
\bqn
\frac{d^{2}x^{\mu}}{ds^2}+\Gamma ^{\mu}_{\nu \rho} \frac{dx^{\nu}}{ds}\frac{dx^{\rho}}{ds}=0,  \lb{geodesics}
\eqn
where $s$ is the affine parameter and $\Gamma ^{\mu}_{\nu \rho}$ are the Christoffel symbols of the quantum-extended Schwarzschild spacetime. For the motions of the stellar stars in the galactic center, it is convenient to consider the weak field approximation. Thus, one can derive the equations of motion of the massive particles by expanding the above geodesics equation in terms of the small quantities, $M$ and $\boldsymbol{v}$, with $\boldsymbol{v}$ being the velocity of the massive particles. In this way, one can transform the above geodesics equation into a perturbed Kepler problem in celestial mechanics, which is described by
\bqn
\frac{d^{2}\boldsymbol{r}}{dt^{2}}&=&-\frac{M}{r^{2}}\frac{\boldsymbol{r}}{r}+\frac{(4+6A_{\lambda})M^{2}}{r^{3}}\frac{\boldsymbol{r}}{r} \nb\\
&& -\frac{M\boldsymbol{v^{2}}}{r^{2}}\frac{\boldsymbol{r}}{r}+\frac{4M}{r^{3}}(\boldsymbol{r}\cdot\boldsymbol{v})\boldsymbol{v}. \lb{geo}
\eqn
Compared to the two-body system described by Newtonian mechanics, 
\begin{equation}
\frac{d^2\boldsymbol{r}}{dt^2} = -\frac{M}{r^2}\frac{\boldsymbol{r}}{r},
\end{equation}
the effective force acting on a massive test particle in the quantum-extended Schwarzschild black hole can be expressed in terms of Newtonian gravitational force plus a perturbated force $\boldsymbol{F}$ as
\begin{equation}
\frac{d^2\boldsymbol{r}}{dt^2} = -\frac{M}{r^2}\frac{\boldsymbol{r}}{r} + \boldsymbol{F}, \lb{EoM1}
\end{equation}
with
\bqn
\boldsymbol{F} &=& \frac{(4+6A_{\lambda})M^{2}}{r^{3}}\frac{\boldsymbol{r}}{r} -\frac{M\boldsymbol{v^{2}}}{r^{2}}\frac{\boldsymbol{r}}{r}+\frac{4M}{r^{3}}(\boldsymbol{r}\cdot\boldsymbol{v})\boldsymbol{v} \lb{FF}
\eqn
which depending on $\boldsymbol{r}$, $\boldsymbol{v}$, and $t$. This perturbed force $\boldsymbol{F}$ contains contributions from both GR and LQG. If $A_\lambda=0$, i.e., when the effect of LQG is absent the three components of $\boldsymbol{F}$ in (\ref{FF}) exactly reduce to those terms arising from GR.

It is well-known that in Newtonian mechanics when the perturbated force $\boldsymbol{F}$ is absent, the bound orbits of a massive object that governs by (\ref{EoM1}) should be a Keplerian elliptical orbit, 
\bqn
r=\frac{a(1-e^2)}{1+e\cos (\theta-\omega)}, \lb{r}
\eqn
where $a$ is the Semi-major axis, $e$ the eccentricity, $\omega$ the argument of pericenter for the elliptic orbit. For such a  Kepler elliptic orbit, one has
\bqn
\dot{r}&=&\sqrt{\frac{M}{p}}e\sin (\theta-\omega),\lb{dotr} \\ 
\dot{\theta}&=&\sqrt{\frac{M}{p^3}}\Big[1+e\cos(\theta-\omega)\Big], \lb{dottheta}
\eqn
where $\dot{r}$ is the radial velocity, $\dot{\theta}$ is the angular velocity, and $p=a(1-e^2)$ is the semi-latus rectum of the elliptic orbit.

To study the effect of the perturbed force $\boldsymbol{F}$, one can project $\boldsymbol{F}$ into three directions $(\boldsymbol{ e_{r}}, \boldsymbol{ e_{\theta}}, \boldsymbol{ e_{z}})$, 
\bqn
\boldsymbol{F}=\mathcal{R}\boldsymbol{ e_{r}}+\mathcal{S}\boldsymbol{e_{\theta}}+\mathcal{W}\boldsymbol{e_{z}},
\eqn
where $\boldsymbol{ e_{r}}$  along $\boldsymbol{r}$, $\boldsymbol{e_{z}}$  along $\boldsymbol{r \times v}$, and $\boldsymbol{e_{\theta}}$ along $\boldsymbol{e_z \times e_r}$. And the three components ${\cal R}$, ${\cal S}$, and ${\cal W}$ are given by
\bqn
\mathcal{R}&=&\frac{(4+6A_{\lambda})M^{2}}{r^{3}}-\frac{M\boldsymbol{v}^{2}}{r^{2}}+\frac{4M}{r^{3}}(\boldsymbol{r}\cdot\boldsymbol{v})\dot{r},\\
\mathcal{S}&=&\frac{4M}{r^{3}}(\boldsymbol{r}\cdot\boldsymbol{v})r\dot{\theta},\\
\mathcal{W}&=&0.
\eqn
Inserting these three components into the Lagrange celestial equation of motion \cite{Peters:1963ux}, one obtains the equations for the evolution of the orbital elements $X^\alpha$ of the form
\bqn
\frac{dX^\alpha}{dt}=Q^\alpha(X^\beta(t), t).
\eqn
For the orbits of stellar stars described by the six elements $(a, e, \iota, \Omega, \omega, f)$, the evolution equations read \cite{Peters:1963ux},
\bqn
\frac{da}{dt}&=&2\sqrt{\frac{a^{3}}{M}}(1-e^{2})^{-\frac{1}{2}}\Big[e\sin f \mathcal{R}+(1+e\cos f)\mathcal{S}\Big],\\
\frac{de}{dt}&=&\sqrt{\frac{a}{M(1-e^{2})}}\Big[\sin f \mathcal{R} \nb\\
&&~~~~ + \frac{2\cos f +e(1+\cos^{2}f) }{1+e\cos f}\mathcal{S}\Big],\\
\frac{d\iota}{dt}&=&\sqrt{\frac{a}{M(1-e^{2})}}\frac{\cos (\omega + f)}{1+e \cos f}\mathcal{W},\\
\frac{d\Omega}{dt}&=&\sqrt{\frac{a}{M(1-e^{2})}}\frac{\sin (\omega +f)}{\sin \iota (1+e\cos f) }\mathcal{W},\\
\frac{d\omega}{dt}&=&\frac{1}{e}\sqrt{\frac{a}{M(1-e^{2})}}\Big[-\cos f \mathcal{R}+\frac{2+e\cos f}{1+e \cos f}\sin f \mathcal{S} \nb\\
&&~~~~ -e\cot \iota \frac{\sin (\omega +f)}{1+e\cos f}\mathcal{W}\Big],\lb{domega}\\
\frac{df}{dt}&=&\sqrt{\frac{M(1-e^{2})}{a^{3}}}\Big(1+e\cos f\Big)^2 \nb\\
&& +\frac{1}{e}\sqrt{\frac{a}{M(1-e^{2})}}\Big[\cos f \mathcal{R}-\frac{2+e\cos f}{1+e\cos f}\sin f\mathcal{S}\Big], \nb\\
\lb{df}
\eqn
where $\iota$ is the orbital inclination, $\Omega$ the longitude of ascending node, $f=(\theta-\omega)$ the true anomaly of the elliptic orbit. By using $\frac{df}{dt}$, one can transform these equations to
\bqn
\frac{da}{df} &\approx&  2\frac{p^{3}}{M}\frac{1}{(1+e\cos f)^{3}}\mathcal{S},\\
\frac{de}{df} &\approx& \frac{p^{2}}{M}\Big[\frac{\sin f}{(1+e \cos f)^2}\mathcal{R} \nb\\
&& ~~~~ +\frac{2\cos f +e(1+\cos^{2}f)}{(1+e\cos f)^{3}}\mathcal{S}\Big],\\
\frac{d \iota}{df}&\approx& \frac{p^2}{M}\frac{\cos (\omega+f)}{(1+e\cos f)^3}\mathcal{W},\\
\frac{d\Omega}{df}&\approx&\frac{p^2}{M}\frac{\sin (\omega+f)}{\sin \iota(1+e\cos f)^3}\mathcal{W},\\
\frac{d\omega}{df}&\approx&\frac{1}{e}\frac{p^2}{M}\Big[-\frac{\cos f}{(1+e\cos f)^2}\mathcal{R}+\frac{2+e\cos f}{(1+e\cos f)^3}\sin f \mathcal{S} \nb\\
&&~~~~ -e\cot \iota \frac{\sin (\omega+f)}{(1+e\cos f)^3}\mathcal{W}\Big].\lb{df}
\eqn

Then the secular changes in the orbital elements $\mu^{\alpha}$ can be calculated via
\bqn
\dot{\mu}^{\alpha}=\frac{1}{T}\int_{0}^{T}\frac{d\mu^{\alpha}}{dt}dt\simeq \frac{1}{T}\int_{0}^{2\pi}\frac{d\mu^{\alpha}}{df}df,
\eqn
where $T=2\pi\sqrt{a^3/M}$ is the period of an elliptical orbit. Performing these integrals for each orbital element, one gets the drift rates of the five orbital elements,
\bqn
\dot{a}&=&0,\\
\dot{e}&=&0,\\
\dot{\iota}&=&0,\\
\dot{\Omega}&=&0,\\
\dot{\omega}&=&\frac{3M^{3/2}(1-A_{\lambda})}{a^{5/2}(1-e^2)}.
\eqn
It is obvious that only the argument of pericenter $\omega$ of the orbit changes over time, which would cause the pericenter precession of the massive object orbiting the quantum-extended Schwarzschild black hole. The precession per orbit reads 
\bqn
\Delta \omega &=&\frac{6\pi M}{a(1-e^2)}\Big(1-A_{\lambda}\Big).\lb{precession}
\eqn
It is evident that when $A_\lambda=0$, the usual Schwarzschild result is recovered. It is interesting to see that $\Delta \omega$ decreases linearly with the LQG parameter $A_\lambda$.

\section{DATA AND DATA ANALYSIS OF THE  17 S-STARS}

\begin{figure*}
 \centering
\includegraphics[width=\textwidth]{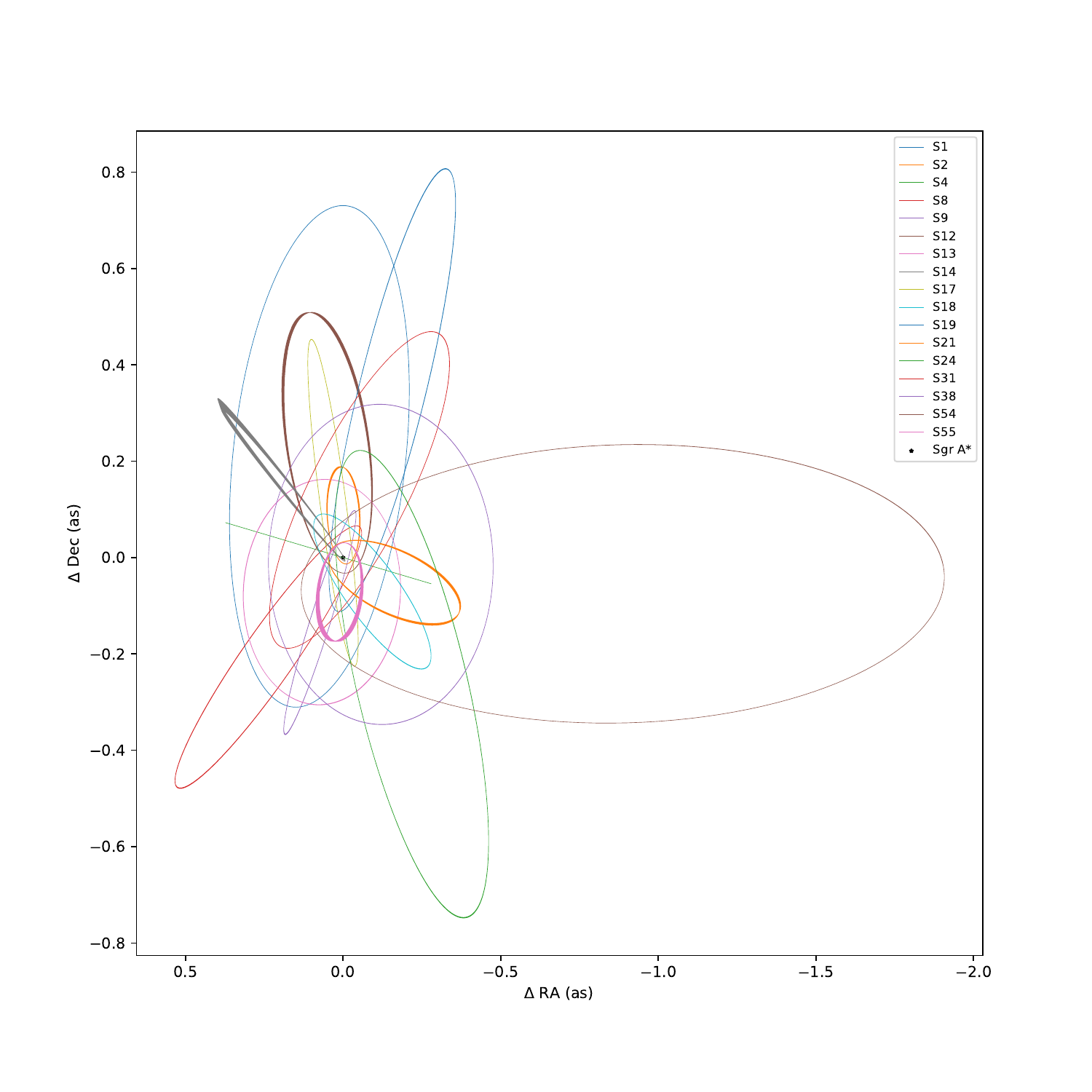}
\caption{The fitted orbits of  the 17 S-stars orbiting around Sgr A* with the best-fit values of orbital parameters from MCMC analysis.} \label{stars}
\end{figure*}

In this section, we present publicly available data \cite{Mon} of the 17 S-stars orbiting around the Sgr A* in our galaxy, and their orbits with the best-fit values of parameters from our MCMC analysis are shown in Figure \ref{stars}. These S-stars include \{S1, S2, S4, S8, S9, S12, S13, S14, S17, S18, S19, S21, S24, S31, S38, S54, S55\}. These data include the data of astrometric positions and radial velocities of 17 S-stars and the pericenter precession of the S2 star. These data are extracted from Ref.~\cite{Mon}.

\subsection{Dataset of positions in the analysis}

The dataset includes the positions of the stars on the sky plane, as well as the corresponding observational times of light $(X_{\rm{data}}(t_{\rm{obs}}), Y_{\rm{data}}(t_{\rm{obs}}))$. By using the Eqs.~(\ref{r}), (\ref{dotr}), (\ref{dottheta}), (\ref{domega}), and (\ref{df}), one can obtain the positions of the stars on the orbital plane and the corresponding emission time of light. Specifically, the positions on the orbital plane are given by $x_{\rm{LQG}}(t_{\rm{em}})=r \cos \theta$ and $y_{\rm{LQG}}(t_{\rm{em}})=r \sin \theta$, where $t_{\rm{em}}$ is the time the light is emitted and $t_{\rm{obs}}$ is the time when the light is observed.

\begin{figure}
 \centering
\includegraphics[width=8.1cm]{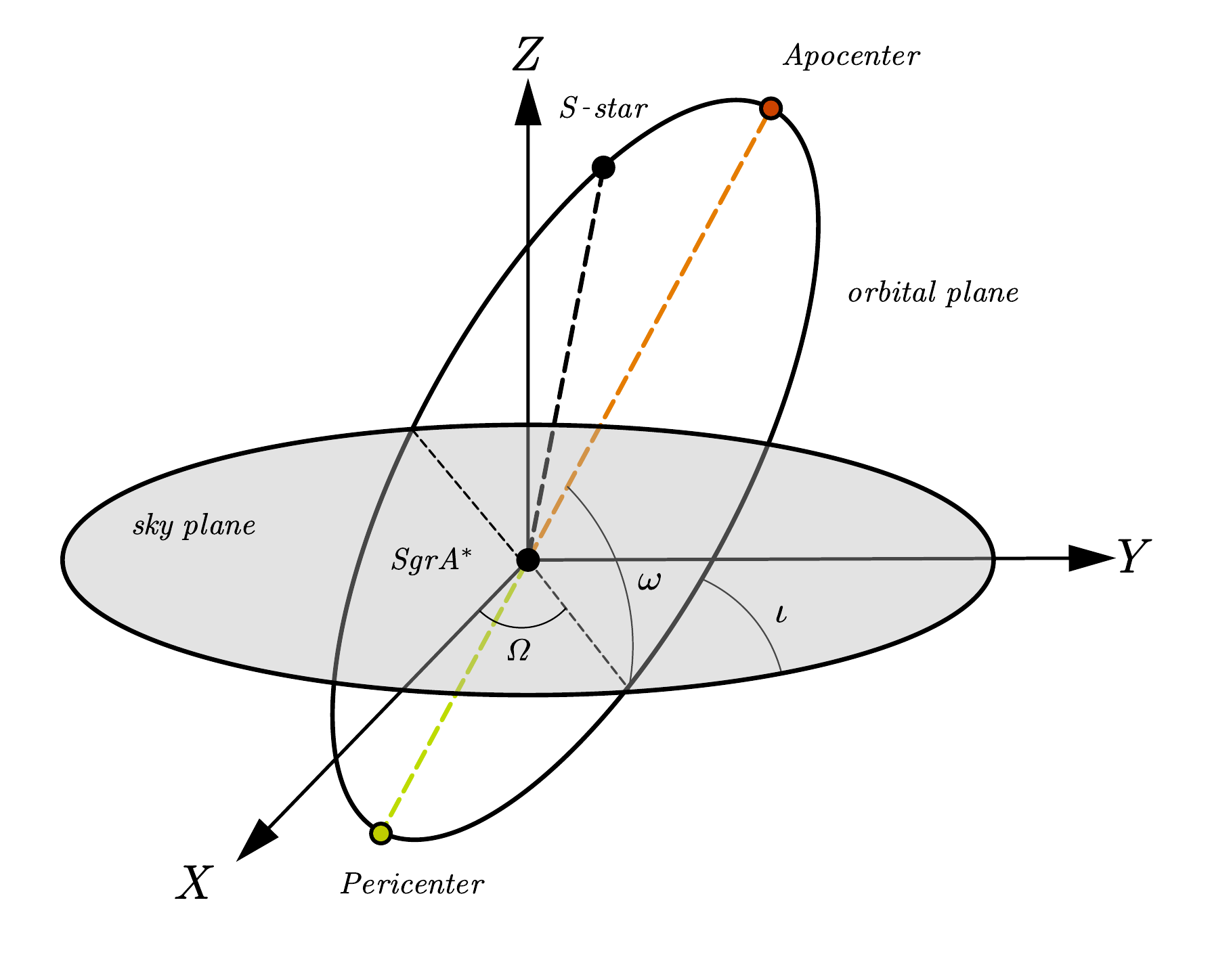}
\caption{This figure illustrates the coordinate system we used for describing the orbits of stellar stars in the galactic center. The $Z$-axis follows the direction the Solar system points to the galactic center and the X-axis points east and the Y-axis points north. $\omega$ is the pericenter argument, $\Omega$ is the longitude of ascending node, and $\iota$ is the orbital inclination of the precessing elliptical orbit for S-stars. } \label{cor}
\end{figure}

To obtain the theoretical positions of the stars, one needs to project their positions in the orbital plane onto the sky plane.  The relation between the sky plane and the orbital plane is illustrated in  Figure \ref{cor}. This involves the Thiele-Innes constants, which relate the star's motion in space to its observed motion in the sky. Specifically, the coordinates on the sky plane $(X, Y, Z)$ are related to the coordinates on the orbital plane $(x, y)$ as follows:
\bqn
X&=&xB+yG, \lb{Xobs}\\
Y&=&xA+yF, \lb{Yobs}\\
Z&=&xC+yH. \lb{Zobs}
\eqn
The Thiele-Innes constants are given by:
\bqn
A&=&\cos \Omega \cos \omega -\sin \Omega \sin \omega \cos \iota, \lb{A1}\\
B&=&\sin \Omega \cos \omega +\cos \Omega \sin\omega \cos \iota, \lb{B}\\
C&=&\sin \omega \sin \iota, \lb{C}\\
F&=&-\cos\Omega \sin \omega -\sin \Omega \cos \omega \cos \iota, \lb{F}\\
G&=&-\sin \Omega \sin \omega +\cos \Omega \cos \omega \cos \iota, \lb{G}\\
H&=&\cos \omega \sin \iota. \lb{H}
\eqn

Next, we need to solve the time delay problem to obtain the correct positions of the stars. The Romer time delay is the main consideration in this work, which can be expressed as:
\bqn
t_{\rm{obs}}-t_{\rm{em}}=\frac{Z(t_{\rm{em}})}{c},
\eqn
where  $c$ is the speed of light. Using the equations (\ref{r}), (\ref{dotr}), (\ref{dottheta}), (\ref{domega}), and (\ref{df}), one can get the positions on the orbital plane and the corresponding emission time of light. Projecting these theoretical positions onto the sky plane using the Thiele-Innes constants (given by equations (\ref{A1}, \ref{B}, \ref{C}, \ref{F}, \ref{G}, \ref{H})) and solving for the Romer time delay, one obtains the projection onto the sky plane of the theoretical positions in the quantum-extended Schwarzschild spacetime $\left(X_{\rm{LQG}}(t_{\rm{obs}}),Y_{\rm{LQG}}(t_{\rm{obs}})\right)$.

\subsection{Dataset of velocity and analysis}

The velocity dataset includes the radial velocity and its corresponding observational time. It is important to consider the photon's frequency shift, denoted as $\zeta$, which affects the radial velocity. This shift can be expressed as
\bqn
\zeta = \frac{\Delta \nu}{\nu} = \frac{\nu_{\rm em}-\nu_{\rm obs}}{\nu_{\rm obs}}=\frac{V_{\rm R}}{c},
\eqn
where $\nu_{\rm em}$ is the frequency of the photon at the time of emission, $\nu_{\rm obs}$ is the frequency when it is observed, and $V_{\rm R}$ is the radial velocity of the S-stars.

In our analysis of the photon's frequency shift, we focus on two relativistic effects: the Doppler shift $\zeta_{\rm D}$ and the gravitational shift $\zeta_G$. These are defined as follows:
\bqn
\zeta_{\rm D}&=&\frac{\sqrt{1-\frac{v_{\rm em}^{2}}{c^{2}}}}{1-\bm n \cdot \bm v_{\rm em}},\\
\zeta_{\rm G}&=&\frac{1}{\sqrt{1 -\frac{2 M}{r \left(\frac{M}{2 r}+1\right)^2}+ \frac{6 A_\lambda M^2}{r^2 \left(\frac{M}{2 r}+1\right)^4} }},
\eqn
where $v_{\rm em}$ is the velocity at the time of emission, and $\bm n \cdot \bm v_{\rm em}$ is the Newtonian line-of-sight velocity. One can then obtain the total frequency shift $\zeta$ as:
\bqn
\zeta= \zeta_{\rm D} \cdot \zeta_{\rm G}-1.
\eqn

\subsection{Orbital precession of S2}
The GRAVITY Collaboration has successfully measured the orbital precession of S2 per orbit, as reported in \cite{GRAVITY2}. The measured value is given by
\bqn
\Delta\phi_{\rm per\;orbit}=12.1\times(1.10 \pm 0.19) .\lb{pre}
\eqn
\red{Different from the measurement of S2's orbit by only using the data of positions and velocities of S2 stars, the measurement of the orbital procession of S2 has used a large amount of new data, as mentioned clearly in \cite{GRAVITY2}. Thus in this paper, we treat this result as an independent measurement and use it independently in our analysis.}
This above orbital precession is an important phenomenon that cannot be explained by the Keplerian orbit under Newtonian gravity, as shown in (\ref{precession}). In addition to the effects of classical gravity, the parameter $A_{\lambda}$ arising from the quantum-extended Schwarzschild spacetime also contributes to the pericenter precession. Therefore, we can use the measured precession of S2 per orbit to constrain $A_{\lambda}$ from an MCMC analysis.

\section{Analysis of Monte Carlo Markov Chain}

In this section, we perform the analysis of MCMC  \red{by open-source \textit{emcee} package in Python} to obtain the constraints of $A_\lambda$ in the quantum-extended Schwarzschild spacetime. We  The parameter space we explored through the MCMC analysis are summarized below
\bqn
\{M, R, a, e, i, \omega, \Omega, T_{\rm P}, x_{0}, y_{0}, v_{x_{0}}, v_{y_{0}}, v_{z_{0}}, A_{\lambda}\},\lb{paras} 
\eqn
 where $M$ is the mass of the central black hole in Sgr A* and $R$ is the distance between the Earth and the black hole, $T_{\rm p}$ refers to the time of the pericenter of the osculating elliptical orbit, which we choose as the starting point of the calculation. $\{ a, e, \iota, \omega, \Omega \}$ are the five orbital elements that describe the osculating elliptical orbits of each S-star. The  parameters $\{x_{0}, y_{0}, v_{x_{0}}, v_{y_{0}}, v_{z_{0}}\}$ represent the zero-point offsets and drifts of the reference frame's center.

To ensure that the parameters are not biased by the choice of priors, we choose to use uniform priors for all parameters. The range of the priors is based on previous results in \cite{Mon}, and we set the prior for $A_{\lambda}$ to be $[0, 2]$. For the likelihood function $\mathcal{L}$, we use three parts: the positions, radial velocities, and orbital precession
\bqn
\log {\cal L} = \log {\cal L}_{\rm P} + \log {\cal L}_{\rm VR} + \log {\cal L}_{\rm pre}.\lb{likelyhood}
\eqn
\red{Since the covariance matrix is not available from the public data we got, we do not consider correlations between data.} Therefore, the likelihood of the positions, $\log {\cal L}_{\rm P}$, is defined as
\begin{equation}
\begin{aligned}
\log {\cal L}_{\rm P} = &-& \frac{1}{2} \sum_{i} \frac{(X_{\rm data}^i -X_{\rm LQG}^i)^2}{(\sigma^i_{X,{\rm data}})^2} \\
&-&\frac{1}{2} \sum_{i} \frac{(Y_{\rm data}^i -Y_{\rm LQG}^i)^2}{(\sigma^i_{Y,{\rm data}})^2},
\end{aligned}
\end{equation}
where $X_{\rm data}^i$ and $Y_{\rm data}^i$ are the measured positions of the star at time $i$, $X_{\rm LQG}^i$ and $Y_{\rm LQG}^i$ are the corresponding positions predicted by our model, and $\sigma^i_{X,{\rm data}}$ and $\sigma^i_{Y,{\rm data}}$ are the uncertainties in the measurements. The likelihood of the radial velocities, $\log {\cal L}_{\rm VR}$, is defined as
\begin{equation}
\log {\cal L}_{\rm VR} = \frac{1}{2} \sum_{i} \frac{(V_{\rm R, data}^i - V_{\rm R, LQG}^i)^2}{(\sigma^i_{V_{\rm R, data}})^2},
\end{equation}
where $V_{\rm R, data}^i$ is the measured radial velocity of the star at time $i$, $V_{\rm R, LQG}^i$ is the corresponding radial velocity predicted by our model, and $\sigma^i_{V_{\rm R, data}}$ is the uncertainty in the measurement. 
The likelihood of the orbital precession, $\log {\cal L}_{\rm pre}$, is defined as
\begin{equation}
\log {\cal L}_{\rm pre} = - \frac{1}{2} \frac{(\Delta \phi_{\rm data}-\Delta \phi_{\rm LQG})^2}{\sigma^2_{\Delta \phi, {\rm data}}},
\end{equation}
where $\Delta \phi_{\rm data}$ is the measured orbital precession of S2 (\ref{pre}), $\Delta \phi_{\rm LQG}$ is the corresponding orbital precession predicted by our model with the given value of $A_{\lambda}$, and $\sigma^2_{\Delta \phi, {\rm data}}$ is the uncertainty in the measurement.

Since we only have the measurement of S2' s orbital precession, $\log {\cal L}_{\rm pre}$ will only be used in the MCMC analysis of S2. By combining these likelihood functions, we can obtain the overall likelihood function $\log {\cal L}$ as defined in (\ref{likelyhood}).

\section{Results}

With all preparations made, we performed the MCMC analysis separately for each of the 17 S-stars to constrain the LQG parameter $A_{\lambda}$ in the quantum-extended Schwarzschild black hole. It is important to note that we only added the orbital precession likelihood to the MCMC analysis of S2, as it is the only star for which we have detected its orbital procession. 

\begin{figure}
 \centering
\includegraphics[width=8.5cm]{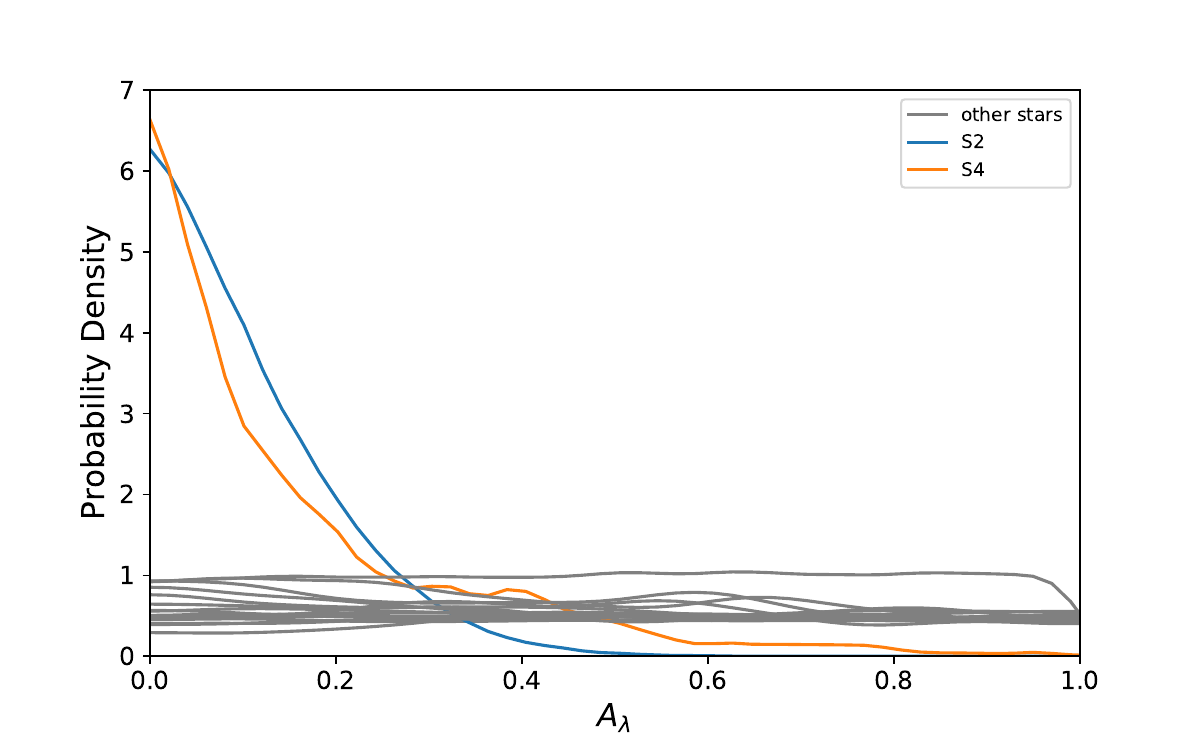}
\caption{The marginalized posterior distributions of the LQG parameter $A_\lambda$ from the analysis with all the 17 S-stars. We performed kernel density estimation(KDE) on these data and show them in the same figure to compare the results. One can easily find out that only the data of S2 and S4 can be used to constrain $A_{\lambda}$ well.} \label{A}
\end{figure}

\begin{figure}
\centering
\includegraphics[width=8.5cm]{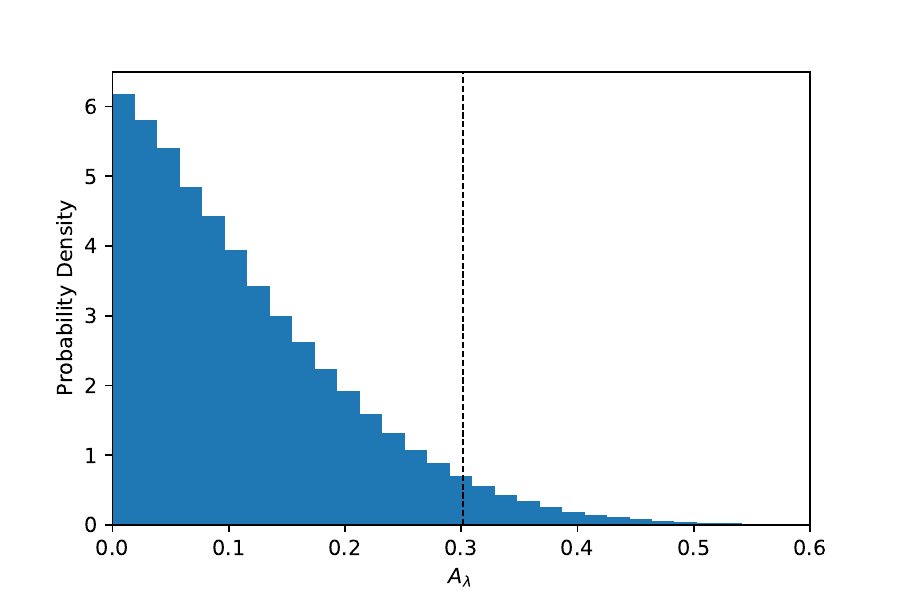}
\caption{The constraint results of $A_{\lambda}$ by the data of S2, $A_{\lambda}<0.302$ at 95\% confidence level. We extract the constraint results of $A_{\lambda}$ from fig \ref{S2} and use 
 a bar chart to show its density distribution which looks like a half Gaussian distribution. One can see the peak value is $0$ which is consistent with the Schwarzschild result. } \label{Alam}
\end{figure}

The results of the marginalized posterior distributions of the LQG parameter $A_\lambda$ from the analysis with all the 17 S-stars are presented in Fig.~\ref{A}. The orange curve represents the result constrained by the data of S2, and the green curve represents the result constrained by the data of S4. We observe that only the data of S2 and S4 can lead to meaningful constraints on $A_\lambda$. By comparing the results from the data of S2 and S4, one observes that the constraint from S2 is stronger than that from S4. Therefore, we select the value constrained by the data of S2 to be the final bound on $A_{\lambda}$, which places an upper limit on $A_\lambda$,
\bqn
A_\lambda \lesssim 0.302
\eqn
at 95\% confidence level. We also plot the marginalized posterior distributions of $A_\lambda$ from S2 in Figure \ref{Alam}. In this figure, the vertical dash line denotes the 90\% upper limits of $A_{\lambda}$.

\begin{figure*}
 \centering
\includegraphics[width=\textwidth]{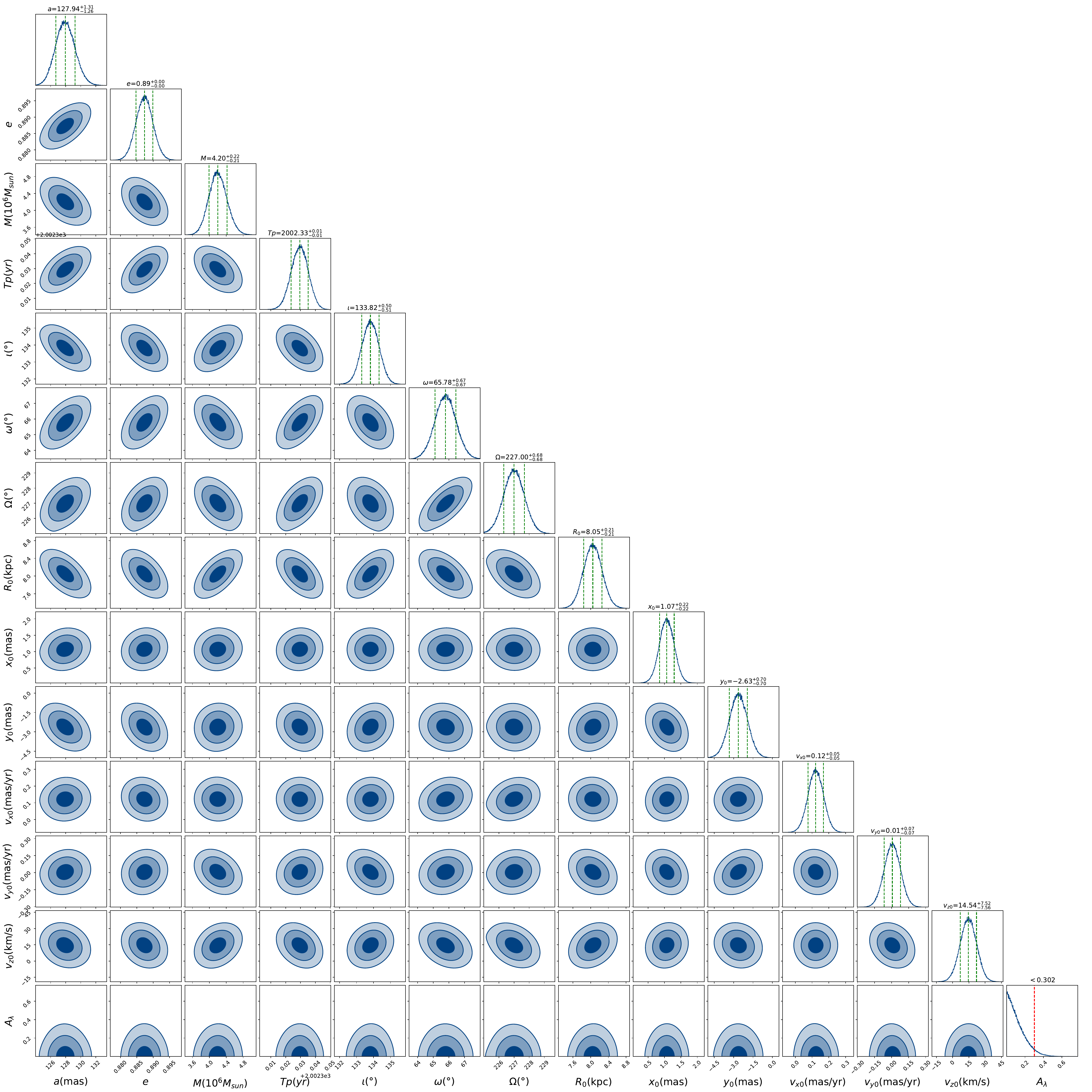}
\caption{The posterior distribution of the orbital parameters of the S2 star and the LQG parameter $A_\lambda$ of the quantum-extended Schwarzschild black hole with uniform priors for all parameters. The LQG parameter $A_\lambda$ is constrained to be $A_{\lambda}<0.302$ at 95\% confidence level. } \label{S2}
\end{figure*}

\begin{figure*}
 \centering
\includegraphics[width=\textwidth]{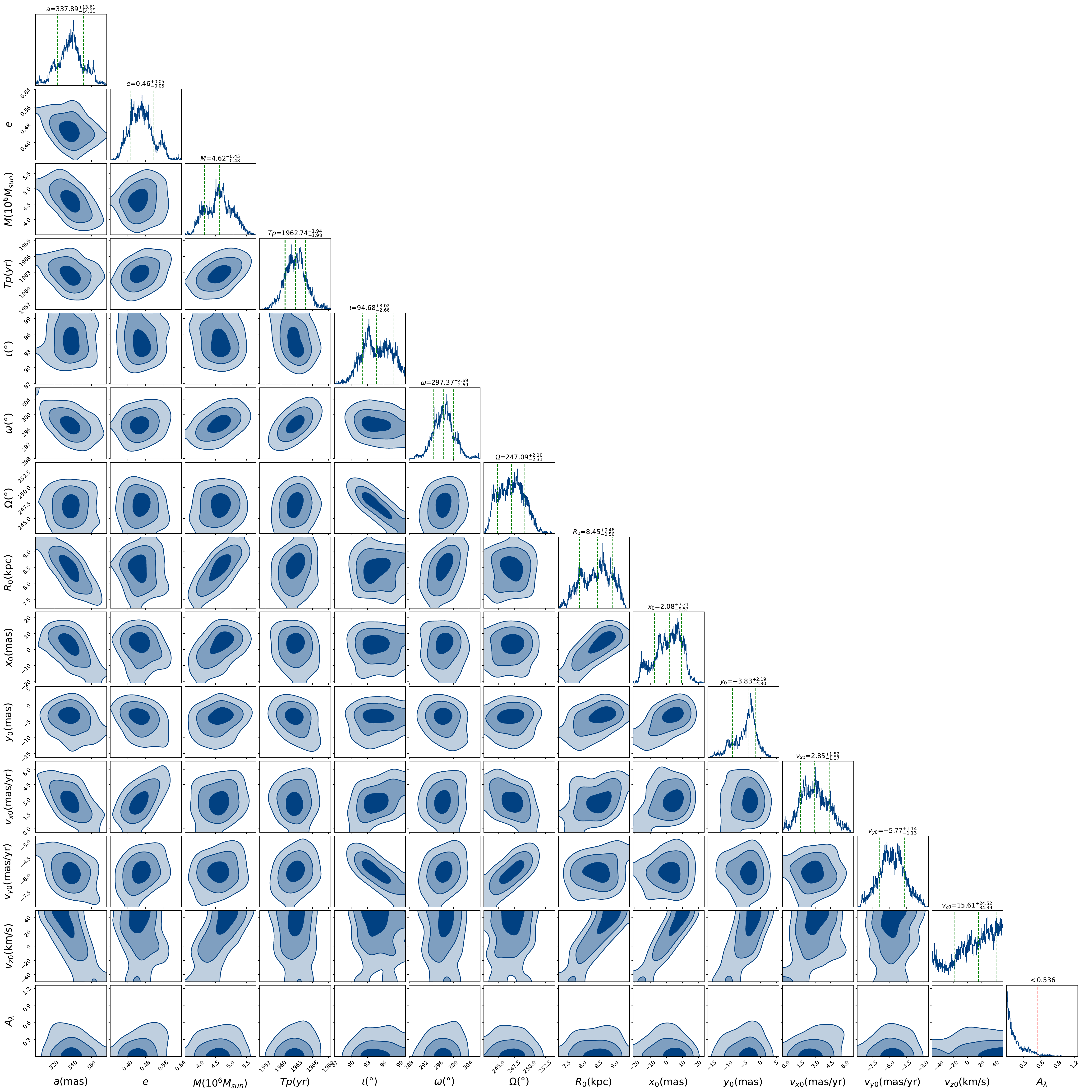}
\caption{The posterior distribution of the orbital parameters of the S4 star and the LQG parameter $A_\lambda$ of the quantum-extended Schwarzschild black hole with uniform priors for all parameters. The LQG parameter $A_\lambda$ is constrained to be $A_{\lambda}<0.536$ at 95\% confidence level.} \label{S4}
\end{figure*}

In Figures \ref{S2} and \ref{S4}, we illustrate the full posterior distributions of our 14-dimensional parameter space of our orbital model for S2 and S4, respectively. On the contour plots of both figures, the shaded regions show the 68\%, 90\%, and 95\% confidence levels of the posterior probability density distributions of the entire set of parameters, respectively.

As expected, the MCMC analysis shows that the data of S2 gives the best result due to its large data size and the inclusion of the measurements of the orbital precession. According to the principle of MCMC, larger data sizes can result in more accurate results, while the inclusion of the orbital precession can help break the degeneracy between the parameters $M, a, e,$ and $A_{\lambda}$ as shown by the precession per orbit function (\ref{precession}). However, it is important to note that the degeneracy could not be completely broken. Figure \ref{S2} displays contour maps that vary from circular to elliptical, indicating the presence of degeneracy between some parameters. \red{In addition, we would like to mention here that in the above analysis, we perform MCMC analysis separately for each of the 17 S-stars. It is interesting to perform a global analysis with all 17 stars together, which is computationally expensive due to a large number of orbital parameters. We expect to come back to this issue in our future works. }

Clearly, the result we obtained is not as strong as those obtained from observations on the scale of the solar system, such as the gravitational time delay measured by the Cassini mission or the perihelion precession of Mercury \cite{Liu:2022qiz}. However, our result demonstrates that observations at the galactic center can provide constraints on black hole parameters beyond those predicted by GR. Our analysis offers a bound on the black hole parameter in the strong gravity regime, which differs from the constraints obtained from observations in our solar system.

\section{conclusion}

In this paper, we introduced the quantum-extended Schwarzschild spacetime and analyzed the motion of massive particles in this spacetime using the dynamics of perturbation. Our findings showed that the orbit of massive particles in this spacetime is a precessing ellipse, and the parameter $A_{\lambda}$ arising from the quantum-extended Schwarzschild spacetime affects the pericenter advance per orbit. To constrain the value of $A_{\lambda}$, we compared the effects of the quantum-extended of the Schwarzschild spacetime with publicly available data of 17 S-stars orbiting around Sgr A* in the central region of the Milky Way. We then carried out an MCMC analysis with all of this data.

To avoid the effects of degeneracy, we gave uniform priors for all 14 parameters. Additionally, based on the fact that S2 has the biggest data size and the most accurate data of precession, we expected S2 to give the most reliable result. As expected, S2 gave the best constraint result, with $A_{\lambda}<0.302$ at the 95\% confidence level. It is also worth mentioning that we ignored the effects of the angular momentum of this spacetime since we expected the effects caused by rotation to be very small.

\section*{Acknowledgements}

This work is supported in part by the Zhejiang Provincial Natural Science Foundation of China under Grant No. LR21A050001 and LY20A050002, the National Key Research and Development Program of China Grant No.2020YFC2201503, the National Natural Science Foundation of China under Grant No. 12275238, No. 11975203, No. 11675143, and the Fundamental Research Funds for the Provincial Universities of Zhejiang in China under Grant No. RF-A2019015.

\end{document}